\documentclass{aa}
\usepackage{amsmath}
\usepackage{natbib}
\usepackage{txfonts}

\ifx\pdftexversion\undefined
\usepackage[dvips]{graphicx}
\usepackage[dvips]{hyperref}

\else
\usepackage[pdftex]{graphicx}
\usepackage{epstopdf}
\usepackage[pdftex,colorlinks=false,pdfstartview=FitV,urlcolor=pdfwww,
  bookmarks=true,bookmarksnumbered=true]{hyperref}

\fi

\bibpunct{(}{)}{;}{a}{}{,} % to follow the A&A style

\newcommand{\deleted}[1]{{\bf\ (DELETED TEXT)}}

\begin{document}

\title {A simulation pipeline for the Planck mission}
\author {M.~Reinecke \inst{1}
\and K.~Dolag \inst{1}
\and R.~Hell \inst{1}
\and M.~Bartelmann \inst{2,1}
\and T.~En\ss lin \inst{1}}
\institute {Max-Planck-Institut f\"ur Astrophysik, Karl-Schwarzschild-Str.~1, 85741 Garching, Germany
\and Institut f\"ur Theoretische Astrophysik, Universit\"at Heidelberg,
Albert-\"Uberle-Str.~2, 69120 Heidelberg, Germany}

\offprints {M.~Reinecke,\\ \email{martin@mpa-garching.mpg.de}}

\abstract {We describe an assembly of numerical tools to model the output data
of the \emph{Planck} satellite \citep{tauber-2004}. These start with
the generation of a CMB sky in a chosen cosmology, add in various foreground
sources, convolve the sky signal with arbitrary, even non-symmetric and
polarised beam patterns, derive the time ordered data streams measured by the
detectors depending on the chosen satellite-scanning strategy, and include
noise signals for the individual detectors and electronic systems.  The
simulation products are needed to develop, verify, optimise, and characterise
the accuracy and performance of all data processing and scientific analysis
steps of the Planck mission, including data handling, data integrity checking,
calibration, map making, physical component separation, and power spectrum
estimation. In addition, the simulations allow detailed studies of the impact
of many stochastic and systematic effects on the scientific results. The
efficient implementation of the simulation allows the build-up of extended
statistics of signal variances and co-variances. \\ Although being developed
specifically for the \emph{Planck} mission, it is expected that the employed
framework as well as most of the simulation tools will be of use for other
experiments and CMB-related science in general.   }

\maketitle

\section {Introduction}

The simulation of realistic data streams is a very important task in
the preparation for the \emph{Planck} satellite mission. Scientifically,
simulated data are required for a realistic assessment of the mission
goals, and for planning how these goals can best be achieved. The
performance of data-analysis algorithms and software needs to be
tested against the scientific requirements of the mission. This
pertains to methods for removing systematic noise from the maps,
separating the multi-frequency data into different physical emission
processes, deriving
power spectra and higher-order statistical measures for CMB
temperature and polarisation fluctuations, identification of ``point''
sources like distant galaxies and galaxy clusters, and many
more. In order to develop, test, and calibrate methods for pursuing
these scientific goals, full-sky maps need to be produced containing
the CMB and all known components of foreground emission in the nine
\emph{Planck} frequency channels between 30 and 857~GHz, at the
resolution of $5'$ and the sensitivity of $\Delta T/T\sim10^{-6}$ to
be achieved by the \emph{Planck} instruments. These maps need to be
``observed'' following the scanning law foreseen for \emph{Planck} and
using realistic beam shapes, filtered with the frequency and time
response of \emph{Planck}'s detectors, and deteriorated by random and
systematic instrumental noise before realistic scientific assessments
can be made.

Also important is the usage of simulated data for the development
of data-analysis techniques for the mission. At a resolution of $5'$,
and assuming sufficient oversampling of the beams, the full sky has on
the order of $5\times10^7$ pixels. Full-sky maps in all frequency bands and
physical components (like foregrounds) need to be efficiently handled and
stored. There
will be 74 detectors on-board \emph{Planck}, 52 on the High-Frequency
Instrument (HFI) and 22 on the Low-Frequency Instrument (LFI). During
one year of the mission, these detectors will produce on the order of
$10^{11}$ measurements along the slowly precessing, stepwise-circular
scanning path, from which the frequency maps need to be
produced. Storing the data efficiently, preparing them for fast
access, and handling them within the data-analysis software pipelines
needs to be thoroughly tested using data simulated at the correct
temporal and spatial sampling rate.

Simulated data are also important for assessing mission operations and
the technical planning. Depending on the scanning law, i.e.~the path
along which the satellite moves and scans the sky, the noise
characteristics of the maps will be more or less anisotropic because
some regions on the sky will be observed much more often than
others. At a fixed scanning law, the position of the detectors in the
focal plane, their orientation relative to the optical
axis, and the orientation of the optical axis relative to the spin
axis of the satellite further affect the scientific goals by changing
the noise characteristics, the coverage pattern on the sky, the
quality of polarisation measurements and more.

Finally, it is an important side-aspect of simulated data that they
can be used to investigate the advantages and disadvantages of
different models for data organisation and storage. With single data
sets exceeding the size of multiple GBytes, efficient data handling
and processing requires that these data be stored in highly
sophisticated ways, e.g.~in data bases such that sequential and random
access to parts of the data is possible. As a European-based mission
with many participating countries, \emph{Planck} data analysis will
operate in a distributed fashion (i.e.\ in several geographically separate
data processing centres), which places further stricter
requirements on the way data are archived and retrieved.

The arguments above demonstrate that the simulation pipeline is
essential for, and an integral part of, both the HFI and LFI Data Processing
Centres \citep{pasian-sygnet-2002}.

Even before the announcement of opportunity for the \emph{Planck} instruments
was addressed, and actually during the phase-A studies for the mission,
several simulations were made to understand the possible science outcome
from \emph{Planck} and to define the instrumental setup. The need was felt
to put on the same footing such simulation effort, and to extend within a
coordinated environment.

These considerations led to the construction of the \emph{Planck}
simulation pipeline which began quite early in the
project, approximately six years ago. At the time there were numerous
simulation activities within many groups involved in the project,
but those simulations were incomplete, aiming at investigating a
multitude of specific and different goals. They were widely used by different
groups throughout the Planck consortia, but produced
data in non-standardised and non-portable formats, and were not
optimised with respect to their consumption of computer resources.

Starting with such existing simulation programs, the \emph{Planck}
simulation pipeline was built according to the following design
guidelines:

\begin{itemize}

\item The pipeline must be modular, i.e.~different simulation tasks
  must be carried out by separate programs, called modules. This
  simplifies extending the pipeline by adding further simulation tasks
  as they become necessary, and it also facilitates testing and
  maintaining the pipeline and its modules.

\item Data must be exchanged between modules in a simple,
  standardised, and machine-independent way. For the pipeline to be
  modular, the input-output interfaces of the modules are crucially
  important. For the interfaces to be simple and platform-independent,
  it was decided that data be exchanged in the FITS\footnote{\href{http://archive.stsci.edu/fits/fits_standard/}{http://archive.stsci.edu/fits/fits\_standard/}}
  format. It is foreseen that most of the data transfer
  between \emph{Planck} simulation and data-analysis modules will
  ultimately be done through a data base, but this does not change the
  principle of exchanging data through simple and platform-independent
  interfaces.

\item Modules can be written in any suitable programming
  language. Since the only contact of a module to its environment is
  established through its interfaces, the internal implementation of
  the module is of no importance. Instead, allowing programmers to
  choose the language they found most suitable or convenient allowed
  the pipeline to be rapidly extended.

\end{itemize}

Following these criteria, the construction of the \emph{Planck}
simulation pipeline proceeded in four steps.

\begin{itemize}

\item First, simulation programs were collected within the
  \emph{Planck} consortia, converted to modules in the sense that they
  were enclosed within suitable interfaces, and augmented by missing
  modules specifically written to complete the pipeline. In that way,
  it was possible within three months to produce the first simulated
  time-ordered data streams from full-sky temperature maps of the CMB
  and several foreground components, which were convolved with
  realistic beams along the scanning path, filtered according to the
  frequency response and appropriately time-sampled, and had random
  noise added.

\item Second, the function of the pipeline was extended by adding
  standard-compliant modules necessary to simulate the observation of
  polarised signal components, point sources, systematic noise,
  extended beam wings, and the CMB dipole, taking the exact motion of
  the spacecraft into account. Massive production of simulated data
  was started (e.g.\ time-ordered data were produced for different detectors,
  scanning strategies and combinations of foregrounds).

\item In a third phase, the \emph{Planck} pipeline was consolidated by
  improving all individual modules in terms of their algorithms and
  their implementation, and adding modules simulating higher-order
  effects like the emission of moving point sources or the detailed
  motion and rotation of the spacecraft. In parallel, a large part of
  the code base of the pipeline was ported from several programming
  languages to C++.

\item Finally, the entire code was optimised in the sense that memory
  consumption was drastically reduced, computation speed was strongly
  increased, OpenMP parallelisation\footnote{\href{http://openmp.org}
  {http://openmp.org}} was added where
  possible, common functions used by more than one module were
  bundled in central libraries, and strict adherence to language
  standards was enforced for compiler-independence.

\end{itemize}

As a result, there is now a simulation pipeline capable of producing
realistic data streams for a full year of observation with individual
\emph{Planck} detectors within a day of CPU time on a computer with
sufficiently large main memory ($\approx$10GB). For
near-realistic parameter choices the entire pipeline can successfully
be run on standard PCs. Due to the strict adherence to language
standards and avoidance of platform-specific libraries, the package
cleanly compiles on four different types of UNIX platforms in both 32-
and 64-bit architectures. Furthermore, it has proven suitable
to run on a computational grid infra\-structure
\citep{taffoni-etal-2005}. The pipeline code is freely available within
the \emph{Planck} consortium.

After giving a brief overview of the \emph{Planck} mission in
Sect.~\ref{overview}, we describe in Sect.~\ref{layout} the layout of the
pipeline, i.e.\ the
decomposition of its simulation tasks into modules, and in
Sect.~\ref{datatypes} the data it produces.
Sections~\ref{commonalities} and \ref{modules} list
the pipeline components common to all modules, and the individual simulation
modules themselves. A summary is presented in Sect.~\ref{summary}.

\section{Summary of \emph{Planck}'s characteristics}
\label{overview}

Here we briefly summarise some characteristics of the \emph{Planck}
mission which are of crucial importance for the design and construction of
the simulation pipeline, and for understanding its function.

\emph{Planck} will observe the microwave sky from a Lissajou orbit
centred on the outer (and unstable) Lagrangian point L2 of the
Sun-Earth/Moon system, $1.5\times10^6\,\mathrm{km}$ from Earth. This
enables the spacecraft to always have the Earth and the Sun at its
back, and the Moon at a small angle relative to its backward
direction. \emph{Planck} will be spin-stabilised, rotating at
1~\emph{r.p.m.}. The optical axis of its Gregorian telescope points at
an almost right angle with the rotation axis, thus scanning the sky
along circles. The rotation axis will be kept fixed for 60~minutes
and then repointed by $2.5$~arc minutes such that
\emph{Planck} progresses as the Earth moves around the
Sun. \emph{Planck} will thus cover the full sky in half a year.
The full mission (after reaching L2) is expected to last 18 months.

The rotation is governed by the inertial tensor of the spacecraft,
which is time-dependent due to fuel and coolant
consumption. Repointing the spacecraft every hour causes precession
which needs to be damped away at the beginning of each pointing
period.

The angle between the optical and rotation axes is slighly less than
$90^\circ$. For \emph{Planck} to cover the full sky, its rotation axis
will make slow excursions from the Ecliptic plane with an
amplitude of several degrees. This scanning pattern implies that the sky
will be covered inhomogeneously, least densely near the Ecliptic and
most densely along lozenge-shaped curves centred on the Ecliptic
poles. Rings on the sky scanned in consecutive pointing periods will
intersect, allowing cross calibration. The scanning pattern also
implies that contact with the receiving antenna in Perth (Australia)
will be limited to two hours per day, during which data down- and
uploads will have to be completed.

Two instruments will receive the incoming microwave radiation. The
low-frequency instrument (LFI) has high electron mobility transistors
(HEMTs) as detectors, allowing measurement of amplitudes and phases of
incoming waves, and thus of their intensity and polarisation. It has
three channels at 30, 44 and 70~GHz. The high-frequency instrument
(HFI) uses bolometers instead. Four of its channels at 100, 143, 217 and
353 GHz are polarisation-sensitive, while the highest-frequency ones at
545 and 857 GHz are not. Passive cooling is insufficient for these
instruments. LFI will operate at 20~K, while HFI needs to be cooled
down to 100~mK. This will be achieved by a four-stage cooling
system. Residual thermal coupling between the cooling system, in
particular the sorption cooler, and the detector electronics, the
instruments and the telescope gives rise to a noise component which
varies with the cyclic operation of the cooler.

The angular resolution of the beams decreases from about 30 to 5 arc
minutes from the 30 to the 857~GHz channels. While the cores of the
beams have approximately Gaussian profiles with slightly elliptical
cross sections, their wings are extended and can receive weak signals
from regions of the sky which are quite far away from the direction of
the beam core. These so-called far-side beam lobes thus add a weak
noise contribution. The feed horns of the individual detectors are
slightly inclined from their positions in the focal plane towards the
optical axis of the telescope, which implies that at a given time all detectors
are observing different points on the sky.

The channel width is approximately 20\% of the central frequency for LFI, and
30\% for HFI. During scans, the signal received by the detectors
will be integrated over certain time intervals and read out at
frequencies varying between 30 and 200~Hz. The sampling frequency needs
to be high enough for the beam to proceed sufficiently less than its
own width along the scanning path during one sampling period.

Slow gain drifts in the detector sensitivity cause \emph{Planck}'s
sensitivity to vary slowly along the circular scan paths. The
frequency power spectrum of these gain drifts falls approximately as a
power law in frequency and is thus called $1/f$ noise. This noise
component gives rise to stripe patterns in the measurements which
follow the scan circles on the sky.

\section {Pipeline layout}
\label{layout}

\begin{figure}
\centerline{\includegraphics[width=.45\textwidth]{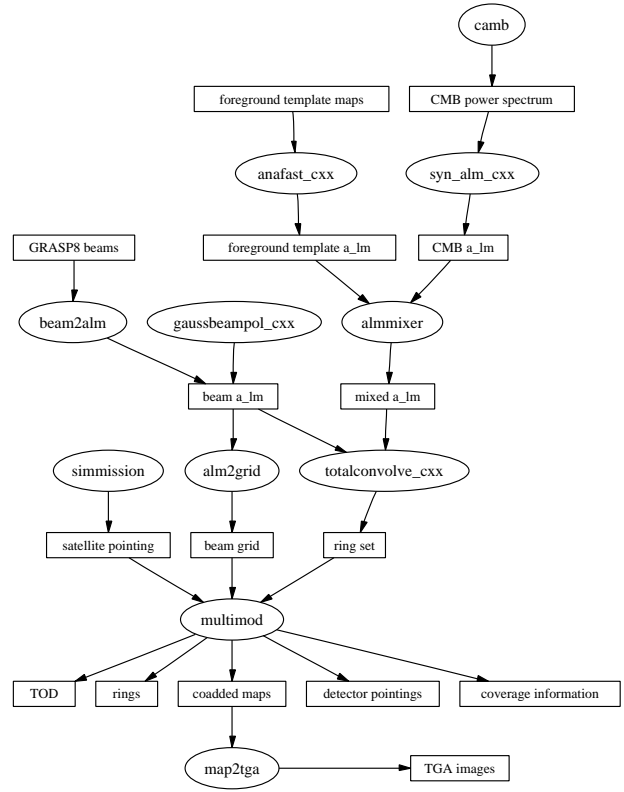}}
\caption{Schematic data flow in a typical \emph{Planck} simulation
  pipeline. Rectangular components denote parameters or data products
  (see Sect.~\ref{datatypes}), whereas elliptic shapes represent modules
  (see Sect.~\ref{modules}). }
\label{dataflow}
\end{figure}

Fig.~\ref{dataflow} illustrates a typical pipeline setup used for
simulating detector timelines and pointings starting from basic inputs
like cosmological parameters, microwave emission models for
galactic and non-galactic components and a scanning strategy.

A short description of the individual steps is given below:
\begin{itemize}
  \item Initially, a CMB power spectrum is generated from a set of
    assumed cosmological parameters, using the CAMB code by Lewis and
    Challinor (see Sect.~\ref{camb}; \citealt{lewis-etal-2000}).
  \item A particular realisation of this power spectrum is generated
    using the {\tt syn\_alm\_cxx} module (Sect.~\ref{healpix}) in
    terms of a set of spherical-harmonic coefficients ($a_{lm}$).
  \item The available set of diffuse foreground component maps is
    transformed into spherical-harmonic coefficients using
    the {\tt anafast\_cxx} module (Sect.~\ref{healpix}).
  \item A combined set of $a_{lm}$ coefficients is generated using the
    {\tt almmixer} module, weighting the individual component maps by
    convolving their electromagnetic spectra with the frequency
    response of the chosen \emph{Planck} detector (Sect.~\ref{mixers}).
  \item A spherical-harmonic transform of an idealised beam template
    is generated using the {\tt gaussbeampol} module; alternatively, a
    realistic, simulated beam pattern can be expanded into spherical
    harmonics using {\tt beam2alm} (Sect.~\ref{beam}).
  \item Since a real-space representation of the beam is required for efficient
    convolution with point-sources, the {\tt alm2grid} code
   (Sect.~\ref{healpix}) is used to generate a beam
    profile sampled on a grid which is equidistant in the polar angles
    ($\vartheta$, $\varphi$).
  \item Using the spherical-harmonic coefficients of the combined sky
    and the beam, the {\tt totalconvolve\_cxx} module produces a
    so-called \emph{ring set} (see Sect.~\ref{datatypes}),
    from which the intensity measured by
    the detector can be determined for all detector pointings and
    orientations (Sect.~\ref{totalconvolve}).
  \item The {\tt simmission} module produces a time stream of
    satellite locations and orientations from input values describing
    the mission orbit parameters, the scanning strategy and the
    dynamical properties of the satellite (Sect.~\ref{simmission}).
    Additionally, {\tt simmission} can generate
    data tables containing the location of the planets as seen from
    the spacecraft, which are used by the point-source convolver.
  \item The {\tt multimod} code (Sect.~\ref{multimod}) processes the
    ring sets, satellite pointing information and beam profiles, and
    generates time streams containing detector pointings and the
    measured signal. In addition to scanning the convolved diffuse sky
    signal (Sect.~\ref{interpolator}), {\tt multimod} also computes
    the signal of the CMB dipole (Sect.~\ref{dipole}), the
    point-source signal (Sect.~\ref{pointsources}), the noise
    generated by the sorption cooler on-board the spacecraft
    (Sect.~\ref{scnoise}), and the instrumental $1/f$-noise
    (Sect.~\ref{oofnoise}). The sampling and integration techniques used for
    the individual detectors (Sect.~\ref{sampler}) are also taken
    into account.

    Mainly for diagnostic purposes, {\tt multimod} can also generate
    co-added sky maps (which are produced by simply adding every simulated
    sample to the pixel containing the corresponding pointing, and subsequent
    averaging) and coverage maps in
    HEALPix\footnote{\href{http://healpix.jpl.nasa.gov}
    {http://healpix.jpl.nasa.gov}} format, which can in turn
    be converted to Targa\footnote{\href{http://astronomy.swin.edu.au/~pbourke/dataformats/tga/}{http://astronomy.swin.edu.au/~pbourke/dataformats/tga/}} image files by the {\tt map2tga} program.
\end{itemize}

It is evident that, in this layout, the pipeline becomes increasingly
mission-specific from beginning to end: while the calculation of the
CMB power spectrum and the creation of a set of $a_{lm}$
is completely independent of the mission setup,
the {\tt almmixer} module already needs information about the
detectors' frequency response. For the total convolution step, the
geometrical beam properties must be known as well, and in order to
produce time-ordered data, a detailed scanning strategy and quantities
like the satellite's inertial tensor must be specified to the {\tt
simmission} module. The {\tt multimod} program additionally requires
the detectors' noise characteristics, sampling frequencies, and
integration time constants.

Conceptually, the simulation process can be subdivided into two
parts: all modules up to and including the totalconvolver and
beam-generation modules produce whole-sky data without time
dependence; the subsequent modules generate time-dependent, localised
output.

It is important to note that the layout described here is by no means
fixed, but can be adjusted to whatever is needed to perform a
particular simulation task. This can be achieved by using the pipeline
editor code, which is part of the \emph{Planck} process coordinator
package.
% \textbf{More specific here: What is allowed to be changed? We
%  need a reference to the ProC!}

The following sections give a more in-depth description of the data types
and codes constituting the simulation pipeline.
While this documentation should suffice to decide in which situations a
given module or data type can be used, it does not give detailed usage
instructions (such as lists of mandatory parameters). This topic is
covered by the document ``Compilation and Usage of the Planck Simulation
Modules'', available in \emph{Planck}'s LiveLink repository or from
the authors, which serves as a detailed manual for practical use of
the simulation package.

\section {Data products}
\label{datatypes}

This section documents the various types of data sets produced by the
\emph{Planck} simulation package which are intended for further use by
the \emph{Planck} Data Processing Centres.

Unless otherwise stated, all data products of the \emph{Planck}
simulation package containing intensities are stored as antenna
temperatures measured in Kelvin. Angles are generally stored as
radians. The simulation pipeline produces maps and pointing information
in ecliptic coordinates.

\begin{description}
  \item {Maps:}\\ The \emph{Planck} simulation package uses maps in
    the HEALPix pixelisation scheme to store full-sky information like
    CMB intensity and polarisation, foreground components, sky
    coverage information and co-added maps. In addition to
    single-component maps, three-component maps are used to store
    polarised data in the form of the Stokes I, Q and U parameters.\\

  \item {Spherical-harmonic coefficients ($a_{lm}$):}\\ A band-limited
    function $f(\vartheta,\varphi)$ with maximum multipole order
    $l_\text{max}$ defined on the sphere can be expressed as a set of
    spherical-harmonic coefficients $a_{lm}$, which is implicitly
    defined by
    \begin{equation}
      f(\vartheta,\varphi)=
      \sum_{l=0}^{l_{\text{max}}}\sum_{m=-l}^l\,
      a_{lm}\,Y_{lm}(\vartheta,\varphi)\;.
    \end{equation}
    These complex coefficients are stored in a three-column format,
    where the first column holds the integer index $l^2+l+m+1$, which
    unambiguously encodes both $l$ and $m$ because $|m|\leq l$. The
    second and third columns contain the real and imaginary parts,
    respectively, of the corresponding coefficient. Since the
    \emph{Planck} simulation package exclusively works on real-valued
    maps, coefficients with negative $m$ are not stored
    because of the condition $a_{l,-m}$ = $a_{lm}^*$ for real-valued
    functions on the sphere. The index column is assumed to contain
    monotonically increasing values, and coefficients not listed in
    the table are assumed to be zero.

    For polarisation information, which can also be stored in the form of
    spherical-harmonic coefficients, the relations between real-space
    and spherical-harmonic space are more involved because the Q and U
    Stokes parameters are quantities of spin $\pm2$, and must
    therefore be expanded in spin-weighted harmonics $_{\pm 2}Y_{lm}$.
    Their exact definition can be found in
    \cite{zaldarriaga-seljak-1997}.\\

  \item {Power spectra ($C_l$):}\\ The power spectra read and
    generated by the \emph{Planck} simulation modules consist of
    either one or four components: both variants contain the
    auto-correlation $C_{l,TT}$ of the temperature fluctuations, the
    four-component variant additionally contains the auto-correlations
    $C_{l,EE}$ and $C_{l,BB}$ of the E and B polarisation modes, as
    well as the cross-correlation $C_{l,TE}$ between temperature and
    E-type polarisation. The following normalisation conventions are
    used:
    \begin{equation}
      C_{l,aa}=\frac{1}{2l+1}\sum_{m=-l}^l\,|a_{lm}|^2
      \qquad\text{(autocorrelation)}
    \end{equation}
    and
    \begin{equation}
      C_{l,ab}=\frac{1}{2l+1}\sum_{m=-l}^l\,a_{lm}^*\,b_{lm}
      \qquad\text{(cross correlation)}
    \end{equation}\\

  \item {Time-ordered data:}\\ The time-dependent output signal of the
    detectors is stored as a single column of data. Since data streams
    for the complete mission can become very large, and many computer
    platforms do not support files larger than 2~GBytes, the
    \emph{Planck} simulation modules can optionally split time-ordered
    data into several files.\\

  \item {Detector pointings:}\\ This time-ordered quantity is stored
    in a two- or three-column format. The first two columns describe
    the co-latitude and the longitude of the beam centre (in radians),
    while the optional third column contains the beam orientation,
    given as the angle between the tangent to the local meridian
    (pointing northwards) and the $x$-axis of the rotated beam
    coordinate system. Similarly to the time-ordered data, detector
    pointings can be split into several files to avoid the 2-GByte
    limit of the file size.\\

  \item {Detector information:}\\ Several \emph{Planck} simulation
    modules need information about \emph{Planck}'s detectors such as
    their position in the focal plane, the orientation of their
    optical axes relative to the telescope's optical axis, their noise
    characteristics, spectral responses, sampling properties,
    ellipticities etc. This information is centrally stored and
    maintained in the so-called \emph{focal plane database}, which was
    originally a text file, but has now been converted to a FITS table
    for usage within the \emph{Planck} simulation package. Recently, a
    new tool was added which allows extracting the required detector
    information from HFI's Instrument Model Object (IMO) files and
    converting it to FITS.

    In addition to the detector characteristics, the focal-plane
    database also contains the angle between the nominal spin axis of the
    satellite and the optical axis of the telescope.\\

  \item {Ring sets:}\\ This data type is written by the {\tt
    totalconvolver} module (Sect.\ \ref{totalconvolve}) and used by
    the {\tt interpolator} module (Sect.\ \ref{interpolator}). It
    contains the three-dimensional array
    \begin{equation}
      T(\phi_E,\phi,m'')=\sum_{m,m'=-l_{\text{max}}}^{l_{\text{max}}}
      T_{m,m',m''}e^{i m \phi_E + i m' \phi}\;,
    \end{equation}
    where $\phi_E$, $\phi$ and $T_{m,m',m''}$ are defined in Eq.~(8) of
    \cite{wandelt-gorski-2001}. We do not carry out the Fourier
    transform over the $m''$ direction because the necessary
    interpolation can be carried out more easily in this
    representation (see Sect.\ \ref{interpolator} for details).

    For real-valued skies and beams, the condition
    $T(\phi_E,\phi,m'')=T^*(\phi_E,\phi,-m'')$ is fulfilled, thus the
    coefficients with negative $m''$ and the imaginary part of
    $T(\phi_E,\phi,0)$ need not be stored.
\end{description}

\section {Common components}
\label{commonalities}

\subsection {External libraries}
\subsubsection {FITS I/O}

Since all modules are presently communicating via FITS files
\citep{fits-1999}, the {\tt cfitsio} library \citep{pence-1999} is
required for the pipeline to operate. Currently, version 2.510 of the
library is used. In the future, it is planned to store the data in a
(possibly object-oriented) data-base system.

\subsubsection {FFT}

Several modules need to perform fast Fourier transformations, which can
be divided into two different usage scenarios:

\begin{itemize}

  \item Most codes (like the transformations of maps between real and
    spherical-harmonic space carried out within the HEALPix package)
    perform FFTs of many different lengths, but each FFT of a given
    length is only performed a few times. Here the
    \emph{Planck} simulation package contains a port to the C language
    of the FFTPACK code originally developed by
    \cite{swarztrauber-1982}. If the length of the required FFT has
    very large prime factors, Bluestein's algorithm
    \citep{bluestein-1968} is employed; the choice of the algorithm
    occurs automatically.

    Both complex- and real-valued transforms are supported.

  \item Several other, still experimental, modules require a highly
    optimised FFT transform of a given length, which is executed many
    times. This task is handled by the {\tt FFTW} library
    \citep{frigo-johnson-1998}, of which the \emph{Planck} simulation
    package currently uses version 3.0.1.

\end {itemize}

\subsection {Other common functions}

Apart from using externally developed general-purpose software such as
{\tt libcfitsio}, the codes have many additional commonalities:
they perform file
input and output in a very similar way (using only a small subset of
the {\tt cfitsio} function), they deal with a common set of data
types (see Sect.~\ref{datatypes}), and they need to operate on them in
similar ways. Keeping a separate implementation of these facilities
for each module would cause unnecessary code duplication, leading to
software that is both error-prone and hard to maintain. For this
reason, common functions are implemented exactly once and linked to
all programs that need it. However, since some modules in the
\emph{Planck} simulation package are written in Fortran 90 and others
in C++, a few common components (most notably the support functions
concerned with data I/O) had to be implemented in both
languages to allow convenient usage.

\subsubsection {Random number generator}

Several modules, most notably the $1/f$-noise generator and the {\tt
syn\_alm\_cxx} module, which creates sets of spherical-harmonic
coefficients from a power spectrum, require streams of random
numbers. Using the default random-number generators supplied with the
programming language is not appropriate in our situation, since the
period length and overall quality of these random numbers depend on
the operating system and the compiler, whereas the data products
created with the \emph{Planck} simulation package should be identical
on all supported platforms (for identical random seeds).

Currently, the \emph{Planck} simulation modules use a generator of the
\emph{xorshift} type \citep{marsaglia-2003} with a period of
$2^{128}-1$, which is more than sufficient for the necessary
calculations. In addition to its excellent randomness properties, the
algorithm is also very efficient.

Since most codes require random numbers with a Gaussian distribution instead
of uniform ones, the simulation package contains an implementation of the
Box-Muller method \citep{box-muller-1958} for generating the former from the
latter.

\subsubsection {Input of simulation parameters}

All \emph{Planck} simulation modules follow a single convention for how
simulation parameters (like cosmological parameters, map resolution,
detector names, names of
input and output files etc.) are obtained from the environment. This
is done via plain ASCII files obeying a simple format. Each parameter
is defined in a separate line using the form:
\begin{verbatim}
  [name] = [value]
\end{verbatim}
Empty lines are skipped. Lines starting with a hash ({\tt \#}) sign are
ignored and can thus be used as comment lines.

\subsubsection {Handling of data I/O}

Data exchange between the various pipeline modules is performed
exclusively via FITS files. Although the {\tt cfitsio} library
provides a comprehensive interface for creating and reading these
files, calling it directly from the modules is in most cases more
complicated and error-prone than necessary for the limited set of data
types used within the \emph{Planck} simulation package. To simplify
data input and output for the module author, a powerful but rather
easy-to-use set of ``wrapper'' functions around {\tt cfitsio} was written
in both Fortran
90 and C++, which supports all input and output operations performed
by the \emph{Planck} simulation modules.

This abstraction from the interface of the underlying library has the
additional advantage that the external data format can be changed with
relatively little impact on module code, simply by adjusting the
implementation (without modifying the user-visible interface)
of the common input-output functions mentioned above to the
new format. The anticipated switch from FITS files to the
\emph{Planck} Data Management Component as a means for transporting
data between modules will greatly benefit from this design.

On an even higher level, functions exist which read and write entire
maps, spherical-harmonic coefficient sets and power spectrum objects;
this code is also used by many modules.

\section {Module descriptions}
\label{modules}

In the following subsections, we describe the function of all
modules in the \emph{Planck} simulation package. Where possible, we
only give a short description of the underlying algorithms and cite
the original works introducing the numerical methods; if no such
publication exists, we provide a more detailed description. For the
most resource-consuming applications we also specify their CPU, memory
and disk-space requirements as well as their scaling behaviour with
regard to relevant input parameters (such as the map resolution or the
maximum multipole order $l_\text{max}$). Parallelisation of a code
is also mentioned where applicable.

\subsection {CAMB}
\label{camb}

For generating CMB power spectra, the November 2004 version of the
CAMB\footnote{\href{http://camb.info}{http://camb.info}} code
\citep{lewis-etal-2000} is used. This code is based on CMBfast
\citep{seljak-zaldarriaga-1996}, but is more portable and can be more
easily integrated into the \emph{Planck} simulation package, mainly
because it does not rely on unformatted files for storage of
intermediate data products.

Memory and CPU requirements of the CAMB code are rather modest for the
parameter sets typically used for \emph{Planck} simulations
(memory consumption\,$<$\,100MB, runtime\,$<$\,1 minute), hence it
can be run without any problems on desktop machines.

\subsection {HEALPix}
\label{healpix}

The \emph{Planck} simulation pipeline makes extensive use of the
HEALPix pixelisation of the sphere
\citep{gorski-etal-2002,gorski-etal-2005}. It is implemented in a
software package which contains many associated algorithms, among
them some exploiting the efficiency of the HEALPix scheme for
spherical-harmonic transforms. The main usage areas of the package
in the simulation pipeline are:
\begin{itemize}
  \item generating spherical-harmonic coefficients from power spectra $C_l$;
  \item conversion between $a_{lm}$ and full-sky maps;
  \item generation of co-added maps from time-ordered data;
  \item generation of efficient look-up tables for the point-source
    convolver.
\end {itemize}

A large part of the HEALPix software package, which is available in
Fortran 90, was re-implemented in C++ in order to be conveniently
accessible from within the C++ simulation modules. During this migration,
the function of the {\tt synfast} tool, which essentially
generates sky maps from power spectra, was split into two separate
facilities, {\tt syn\_alm\_cxx} (which generates $a_{lm}$
from a power spectrum) and {\tt alm2map\_cxx}
(which transforms $a_{lm}$ coefficients to a HEALPix map). As can be
seen in Fig. \ref{dataflow}, this allows feeding the results of {\tt
syn\_alm\_cxx} directly into the {\tt almmixer} module (Sect.\
\ref{mixers}), and avoids the conversion to real space and back to
$a_{lm}$, which was necessary in earlier versions of the \emph{Planck}
simulation package and caused some avoidable loss of accuracy and
waste of computer resources.

The ported code (especially the conversions between maps and $a_{lm}$
coefficients) was optimised with regard to memory and CPU time
consumption. On shared-memory computers, OpenMP parallelisation is
used.

An addition to the HEALPix facilities is the {\tt alm2grid}
program. This code converts $a_{lm}$ coefficients to values on an
equidistant grid in the polar angles ($\vartheta$,$\varphi$), whose
extent in $\vartheta$ can be chosen freely. Using this format, it is
easy to calculate a high-resolution real-space representation of a
beam pattern, which in turn is a necessary input for the point-source
convolver.

The C++ package also contains a testsuite to validate the correctness and
accuracy of the ported code.

Starting with version 2.0, the official HEALPix package contains the
described C++ modules, and is distributed under the GNU General Public
License\footnote{\href{http://www.gnu.org}{http://www.gnu.org}}.

\subsection {skymixer/almmixer}
\label {mixers}

The {\tt skymixer} component serves two purposes:

\begin{itemize}

  \item It converts maps of the CMB and diffuse foreground sources
    (such as the Galactic synchrotron and dust emission and the
    thermal and kinetic Sunyaev-Zel'dovich effects) from the units in
    which they are typically provided (e.g.\ MJy/sr at a certain
    frequency for the Galactic maps, and the Compton
    $y$-parameter for the thermal Sunyaev-Zel'dovich maps) to antenna
    temperature observed by a given \emph{Planck} detector.  During
    this conversion, the frequency response of \emph{Planck}'s
    detectors is taken into account, i.e.\ the frequency-dependent sky
    signal is convolved with the sensitivity of the detectors not only
    at the nominal detector frequency, but in a frequency band around
    it. Currently, the spectral detector response can have Gaussian,
    top hat or $\delta$ form.

  \item It adds the resulting antenna-temperature maps to a single map
    with all selected components.

\end{itemize}

The currently employed code is a re-implementation of a version by
R.~Ansari, which was based on the SOPHYA\footnote{\href{http://sophya.org}{http://sophya.org}} library.

The microwave sources currently supported by {\tt skymixer} are:

\begin{itemize}

  \item CMB given as temperature fluctuations in Kelvin:\\ This kind
    of map is typically generated from a power spectrum $C_l$ produced
    by the CAMB code.

  \item Galactic synchrotron emission, given in MJy/sr at 408\,MHz:\\
    The template was provided by G.~Giardino; details concerning its
    creation are given by \cite{giardino-etal-2002}. This component is
    extrapolated to the \emph{Planck} frequency bands by using a
    power-law spectrum with an exponent of \mbox{-0.75} between
    408\,MHz and 22\,GHz, and with an exponent of \mbox{-1.25} above
    22\,GHz.

    A spectral-index map derived from Wilkinson-MAP data is also
    available, but has not yet been included into the extrapolation
    algorithm.

  \item Galactic dust emission, given in MJy/sr at 100$\mu$m:\\
    Currently this source is handled by a one-component approximation
    provided by C.~Baccigalupi. The full two-component model presented
    by \citet{finkbeiner-etal-1999} is currently being implemented.

    The employed template map was published by
    \cite{schlegel-etal-1998}.

  \item Galactic free-free emission: \\ This map was derived from a
    template map of galactic H$_{\alpha}$-emission published by
    \cite{finkbeiner-2003}, according to the model of
    \cite{valls-gabaud-1998}.

  \item thermal and kinetic SZ-effect, given as the Compton-$y$ and
    Compton-$w$ parameter, respectively:\\ Briefly, these maps were
    produced using the galaxy-cluster distribution in space, peculiar
    velocity and redshift found in the Hubble Volume simulation
    \citep{jenkins-etal-2001,colberg-etal-2000},
    attaching appropriately scaled gas-dynamical cluster
    simulations to the cluster positions and projecting the
    Compton-$y$ and $w$ parameters on the sphere. Details about the
    template maps are given by \cite{schaefer-etal-2004a}.\\
    Additional maps for the thermal and kinetic SZ-effect of the local super
    cluster structure have become available recently. In this case,
    the full sky SZ maps are directly calculated from a hydodynamical,
    constrained simulation of the local universe \citep{dolag-etal-2005},
    which spans a sphere of 110Mpc radius. These maps are complemented by
    considering the galaxy-cluster distribution outside this sphere using the
    Hubble Volume simulation as described before.

  \item Emission by rotating CO molecules:\\ This foreground component
    only contributes to the HFI channels. The template map was
    published by \cite{dame-etal-1996,dame-etal-2001}. We use the
    method given by \cite{schaefer-etal-2004b} to determine the
    intensities in \emph{Planck}'s frequency bands.

\end{itemize}

Since the output of the {\tt skymixer} is most frequently used in a
subsequent run of the {\tt totalconvolve\_cxx} module (see
Sect.~\ref{totalconvolve}) which requires $a_{lm}$ coefficients as input,
and since the synthesised CMB signal can be very efficiently created
in the form of spherical-harmonic coefficients $a_{lm}$, an additional
variant of the {\tt skymixer} code was implemented. This code (named
{\tt almmixer}) performs exactly the same task as {\tt skymixer}, but
operates entirely in spherical-harmonic space. Using this code
eliminates the need for converting between angular and
spherical-harmonic space twice during a run of a typical simulation
pipeline and thus saves a considerable amount of CPU time.

Both codes allow the suppression of the CMB monopole in the output,
which is usually not needed by any subsequent analysis, and
which would decrease the numerical accuracy of the output.

The runtime of both programs is rather short and dominated by the
input and output operations. The memory consumption
of the {\tt skymixer} is low (below 10~MBytes) and independent of the
map size, since it processes them chunk by chunk. In contrast, the
{\tt almmixer} module holds two $a_{lm}$ sets in memory, so that it
requires approximately $16\,l_{\text{max}}^2$ bytes of storage.

\subsection {Beam generation}
\label{beam}

Several \emph{Planck} simulation modules require information about the
beam shapes of \emph{Planck}'s detectors, typically in the form of a
set of spherical-harmonic coefficients.

One way to obtain these is the module {\tt gaussbeampol}, which takes
a detector name as input, looks up the detector's beam characteristics
in the focal-plane database, and produces the corresponding $a_{lm}$
coefficients. The {\tt gaussbeampol} module can generate axisymmetric
or elliptical Gaussian beams with and without polarisation, and allows
arbitrary orientations for polarisation and ellipticity.

Realistic beam patterns are often delivered in the GRASP8 format\footnote{\href{http://www.ticra.com}{http://www.ticra.com}},
which in turn consists of the GRD and CUT sub-formats. The {\tt
beam2alm} code can convert beam files of both sub-formats to $a_{lm}$
coefficients, and also allows combining two GRASP8 files (containing
the full and main beam, respectively) to a single $a_{lm}$ set.

The point-source convolver requires a real-space representation of the
beam, which can be obtained from the $a_{lm}$ coefficients using the
{\tt alm2grid} or {\tt alm2map\_cxx} modules.

\subsection {Total convolution}
\label{totalconvolve}

This module takes as input the spherical-harmonic coefficients of a
sky map and a beam, and computes the convolution of sky and beam for
all possible directions ($\vartheta$, $\varphi$) and orientations
($\psi$) of the beam relative to the sky. Both unpolarised and
polarised convolutions are supported. In the polarised case, three
sets of spherical harmonics are required for sky and beam,
respectively, for their Stokes-I, Q, and U parameters. The output then
consists of the sum of the three separate convolutions.

Our implementation is analogous to the algorithm presented by
\citet{wandelt-gorski-2001}, including the extensions for polarisation
given by \citet{challinor-etal-2000}. However, the code was modified
to allow complete shared-memory parallelisation of all CPU-intensive
tasks (starting from the partial parallelisation contributed by
Stephane Colombi), and the symmetries inherent in the matrix $d_{mm'}^{l}$
(introduced in Eq.~(6) of \citealt{wandelt-gorski-2001}) were used for
further CPU and memory savings. With the new code, convolutions up to
$l_{\text{max}}=1500$ and $m_{\text{max}}=2$ can be done in less than
five minutes on a modern desktop computer.

The resource consumption of the {\tt totalconvolve\_cxx} module scales
roughly with $l_{\text{max}}^2\,m_{\text{max}}$ for memory and
$l_{\text{max}}^3\,m_{\text{max}}$ for CPU time. Although the code is
fully parallelised, it will most likely not scale very well, since the
cache re-use of the algorithm is poor and a saturation of the memory
bandwidth will limit overall performance.

The output consists of the three-dimensional complex-valued array
$T(\vartheta,\varphi,m)$, which contains $l_{\text{max}}+1+x$ points
in $\vartheta$-direction ($x$ is a small number of additional points
needed for interpolation purposes), $2l_{\text{max}}+1$ points in
$\varphi$-direction, and $m_{\text{max}}+1$ values in
$m$-direction. In contrast to the original implementation by
B.~Wandelt, the final FFT is only carried out over $\vartheta$- and
$\varphi$-directions, because interpolation is easier for this case;
see also Sect.\ \ref{interpolator}.

To increase the spatial resolution of the output ring set, the working
array can be zero-padded to any desired $l_{\text{max,out}}\geq
l_{\text{max}}$ before the FFT is performed.

\subsection {Satellite dynamics}
\label{simmission}

The \emph{Planck} simulation package makes use of the {\tt simmission}
code developed by Floor van Leeuwen and Daniel Mortlock
\citep{vanLeeuwen-etal-2002,challinor-etal-2002}, which takes as input
a wide variety of mission-related parameters like the exact orbit
around the outer Lagrangian point L2 of the Sun-Earth/Moon system, the
scanning strategy, the satellite's inertial tensor etc.

The most important output of the {\tt simmission} module is a table
containing the position and orientation of the satellite in fixed time
intervals during the entire mission. In addition, it optionally
calculates the positions of the Sun and planets relative to the
\emph{Planck} spacecraft, which is important for the point-source
convolver.

\subsection {multimod}
\label{multimod}

The {\tt multimod} code comprises most of the functions offered by
the \emph{Planck} simulation package for creating and manipulating
time-ordered information. This includes creation of detector pointings
and signal time streams from various sources like the CMB dipole,
point sources and detector noise.

At first glance, this approach appears to contradict the goal of
modularity stated at the beginning of this paper, which would suggest
writing many small pieces of code implementing a single task each.
This strategy, however, would have a negative impact on performance,
since in most of the simulation codes working on time-ordered data,
the disk input or output times are at least comparable to the
computations themselves. Combining all the effects in one program
allows to read and write the data only once and perform all
manipulations without storing intermediate data products on disk.

Even though all the functions are linked together into one
executable, the various components are still implemented in a very
modular fashion, so that they can be easily used to build smaller modules with
reduced generality. For example, the \emph{Planck} simulation package
contains a stand-alone point-source convolution code, which includes
exactly the same C++ classes also used by the {\tt multimod} code.

The {\tt multimod} program processes data in chunks of one pointing
period at a time, which corresponds to an hour of measurement. This keeps
the amount of main memory occupied by time-ordered data rather low. On
the other hand, a pointing period still contains enough individual
samples to allow efficient OpenMP-parallelisation of the code.

Schematically, {\tt multimod} performs the following steps for each pointing
period:
\begin{enumerate}
\item Calculate the detector pointings at a sampling rate of {\tt ringres}/60s
  from the satellite pointings (see Sect.~\ref{detpoint}).
\item Calculate and add all of the requested signal contributions
  (except for $1/f$-noise) at a sampling rate of {\tt ringres}/60s
  (see Sections \ref{interpolator} -- \ref{scnoise}).
\item Process the calculated signal in the sampler module
  (see Sect.~\ref{sampler}).
\item Add the $1/f$-noise if requested (see Sect.~\ref{oofnoise}).
\item Produce the detector pointings at the detector sampling rate.
  \label{step5}
\item Write the requested signal and pointing streams to disk. \label{step6}
\item Update the coverage and coadded sky maps.
\end{enumerate}
Each of these steps is only executed if its output was requested by the user
or if it is required by subsequent steps; i.e.\ if the user only asks for
the generation of detector pointings, only steps \ref{step5}  and
(partially) \ref{step6} are executed.

The parameter {\tt ringres} can be chosen freely by the user; it determines
the sampling rate used for producing the ``ideal'' detector signal
(i.e.\ the signal before the modification by the detector electronics and
the sampling process). In theory this sampling rate should be infinitely
high; for practical purposes it should be sufficient to choose
{\tt ringres}/60s $\approx\! f_{\text{samp}}$ for smooth signals (like the CMB
and galactic foregrounds) and {\tt ringres}/60s $\approx\! 3f_{\text{samp}}$
for the SZ and point-source signals.

\subsubsection {Detector pointings}
\label{detpoint}

Using the time stream $M_n$ of satellite orientations generated by the
{\tt simmission} module, the detector-pointing component produces
time-ordered detector pointings at arbitrary sampling rates. For this
purpose, it is necessary to determine the satellite orientation at any
given point in time. This is achieved by extracting the rotation
matrix $R_n$ between every $M_n$ and $M_{n+1}$. From this matrix, an
axis $\vec r_n$ and angle $\alpha_n$ of rotation can be computed. The
satellite orientation at a time $t$ with $t_n\leq t<t_{n+1}$ can then
be approximated by
\begin{equation}
  M_t=RM\left(\vec r_n,\frac{t-t_n}{t_{n+1}-t_n}\alpha_n\right)\,M_n\;,
\end{equation}
where $RM(\vec r,\alpha)$ is a rotation matrix performing a rotation
of $\alpha$ around the axis $\vec r$.

After this rotation interpolation another rotation matrix must be
applied, which describes the detector position and orientation relative to
the satellite coordinate frame.
This matrix depends on the angle between the optical axis and the nominal
spin axis, as well as the detector-specific
angles $\phi_{uv}$, $\vartheta_{uv}$ and $\psi_{uv}$ given in the
focal-plane database,
which specify the position and orientation of all detector
horns relative to the coordinate frame of the focal plane.

\subsubsection {Interpolator}
\label{interpolator}

The {\tt interpolator} module's task is to determine the radiation
intensity falling onto a detector for a given co-latitude $\vartheta$,
longitude $\varphi$ and beam orientation $\psi$. To accomplish this,
the output of the {\tt totalconvolve\_cxx} module for the desired sky
signal and the detector's beam is used. As \cite{wandelt-gorski-2001}
point out, this dataset is sufficient (for a given $l_{\text{max}}$
and $m_{\text{max}}$ of the beam) to reconstruct the exact signal for
any ($\vartheta$, $\varphi$, $\psi$)-triple. However, exact evaluation
is prohibitively expensive in practice, since it would involve a sum
over $l_{\text{max}}^2\,m_{\text{max}}$ terms for a single data point.

The approach used in the {\tt interpolator} is based on the assumption
that $l_{\text{max}}$ will be rather large ($\approx 4000$) for
realistic simulations, while the beam's $m_{\text{max}}$ is usually
not larger than 20 (i.e. the beam pattern has no high-frequency
azimuthal asymmetries). The high $l_{\text{max}}$ implies a fairly
high resolution in $\vartheta$ and $\varphi$ for the totalconvolver's
output, so that polynomial interpolation (whose order can be chosen
via a parameter) can be used for those two directions. In the
$\psi$-direction, this approach is not accurate enough because of the
sparse sampling, so that the Fourier sum of the $\psi$-components must
be calculated:
\begin{equation}
  S(\vartheta,\varphi,\psi)=\sum_{m=-m_{\text{max}}}^{m_{\text{max}}}
  T_{\text{interpol}}(\vartheta,\varphi,m)\,\mathrm{e}^{im\psi}
\end{equation}

A straightforward calculation of $\mathrm{e}^{im\psi}$ would be quite
CPU-intensive because of the required trigonometric function calls,
but making use of the trivial recursion relation $e^{i(m+1)\psi}$ =
$e^{im\psi}$\,$e^{i\psi}$ reduces calculation time significantly. This
optimisation is only applicable here because $m_{\text{max}}$ cannot
become very large; if it did, the roundoff errors accumulated in the
recursion could destroy the accuracy of the result.

The memory consumption of the module is proportional to
$l_{\text{max}}^2\,m_{\text{max}}$, its runtime scales with
$m_{\text{max}}\,n_{\text{samples}}$. Since interpolations for
different pointings are independent, parallelisation was trivial.

By default, the {\tt interpolator} module calculates the total
intensity observed by the detector. Optionally, any of the individual
Stokes parameters I, Q and U can also be extracted, but this
calculation is only exact for axisymmetric polarised beams.

When using linear interpolation, it has been observed that the power
spectrum of the resulting co-added maps was suppressed in comparison to
the input maps at high $l$. This is a consequence of the finite
resolution of the ring set, combined with the non-perfect
interpolation in $\vartheta$ and $\varphi$. Although this effect
cannot be eliminated entirely, increasing the interpolation order
and/or increasing $l_{\text{max,out}}$ in the {\tt totalconcolver}
reduces the problem dramatically.

\subsubsection {Dipole}
\label{dipole}

This component calculates the time-dependent dipole signal observed by
a given detector. The user can choose whether the kinematic dipole
component (caused by the motion of the satellite in the Solar System,
with the Solar System within the Galaxy, with the Galaxy within the
Local Group, and so forth) should be included, whether relativistic
effects should be calculated, whether only higher order corrections
should be returned etc. The default output quantity is the dipole
signal in antenna temperature, but thermodynamic temperature can also
be calculated.

\subsubsection {Point-source convolver}
\label{pointsources}

The task of the point source convolver is to simulate time streams of the
signal created by fixed and moving point sources. For fixed sources, this
could in principle be achieved also by adding another foreground to
{\tt skymixer} or {\tt almmixer}, but this is not very practical because
the maximum multipole moment required in the {\tt totalconvoler} to accurately
treat this kind of strongly localised sources would be very high.
For moving sources, this solution is not viable at all, since all signal
contributions entering the mixing modules must be constant over the mission
time scale.

For these reasons, the point source convolver operates in real space, taking
as input the detector pointings and convolving the sources close to a given
pointing with the properly oriented beam pattern. Obviously, an efficient
retrieval of sources in the vicintity of the beam is crucial for the
performance of the module. This was achieved by presorting the point source
catalog into a HEALPix map with low resulution ($N_{\text{side}}$=16), and
by identifying, for all detector pointings, the nearby point sources using
the {\tt query\_disc()} function of HEALPix with a radius of several times
the beam FWHM.

While fixed point sources stay in this HEALPix lookup table during a whole
simulation run, the positions of the moving point sources are calculated once
per pointing period. These positions are then sorted into the lookup table
accordingly and taken out again after the signal for the pointing period has
been calculated.

As mentioned above, it is recommended to choose a high value for the
{\tt ringres} parameter when calculating the point source signal, because
of their very localised nature.

For the fixed point sources, the radiation flux in each \emph{Planck} frequency
band is taken from the point source catalog. For moving point sources
(i.e.\ planets and asteroids), the corresponding positions in time are
calculated using the JPL HORIZONS system\footnote{\href{http://ssd.jpl.nasa.gov/horizons.html}{http://ssd.jpl.nasa.gov/horizons.html}}, whereas the radiation intensity of
the moving point sources is calculated using an extended Wright \& Odenwald
thermal emission model \citep{wright-1976,neugebauer-etal-1971}
which has been adapted separately for rocky planets,
gaseous planets and asteroids, and which
accounts for effects like distance to the satellite, illumination
geometry, beaming, surface roughness, heating and cooling of the surface, etc.
This will be described in more detail in a separate publication
(R.\ Hell, in preparation; see also \citealt{schaefer-etal-2004b}).

\subsubsection {Sorption-cooler noise}
\label{scnoise}

Temperature fluctuations in the detectors caused by the operation of
\emph{Planck}'s sorption cooler constitute an important systematic
effect and have to be included in realistic simulations of
time-ordered data.  To this end, the \emph{Planck} simulation package
contains the {\tt glissando} code contributed by LFI, which models
this effect and will be fully integrated with {\tt multimod}
in the future. Since the LFI and HFI detectors may react quite
differently to temperature fluctuations, it is likely that a
completely different code will ultimately be required for the HFI
channels.

\subsubsection {Sampler}
\label{sampler}

The {\tt sampler} component takes as input the idealised signal
impinging on the detector, which is produced by the components
described above, and applies to it a model of the detector's
measurement process. Two effects play an important role:
\begin{itemize}
  \item Every sample recorded by any detector is in reality the
    average of several \emph{fast samples} taken in the time since the
    last slow sample.
  \item All HFI detectors have an exponential impulse response with a
    time constant $\tau_{\text{bol}}$, so that each fast sample must
    in turn be calculated via the integral
    \begin{equation}
      s_{\text{bol}}(t)=\int_{-\infty}^t dt' s_{\text{sky}}(t')
      \frac{\exp [-(t-t')/\tau_{\text{bol}}]}{\tau_{\text{bol}}}\;.
    \end{equation}
    The algorithm approximates this integral by taking $N$ unequally
    spaced samples:
    \begin{equation}
     s_{\text{bol}}(t)\approx\sum_{k=1}^{N}\,s_{\text{sky}}(t_k')\;,
    \end{equation}
    where $t_k'=t+\tau_{\text{bol}}\ln(k/N)$.
\end{itemize}
A more detailed description of the implementation can be found in
\cite{grivell-mann-1999}.

It should be mentioned that both effects introduce a time delay into
the sampled time-ordered data, i.e.\ that the signal written with the
time stamp $t_n$ was produced entirely from measurements at $t<t_n$,
which in the case of HFI reflect a signal that was seen even earlier,
because of the bolometer's time response. This must be taken into
account when time-ordered signal data and detector pointings are used
to produce sky maps.

The sampler code can be trivially parallelised and has perfect cache
behaviour, so that it should scale very well on multiprocessor
machines.

As expected, the {\tt sampler} module produces data at a rate equal to
the detector's sampling frequency.

\subsubsection {$1/f$-noise}
\label{oofnoise}

The electronics of both LFI and HFI detectors produces noise which has
a power-law spectrum with spectral index $0\le \gamma \le2$ between a
lower frequency $f_{\rm min}$ and a higher (or knee) frequency $f_{\rm
knee}$. Below and above those frequencies, the spectrum is white.
The targeted spectral power is thus
\begin{equation}
\label{eq:targetnoise}
  P_n^*(f)=P_0\,\left(1+ (f/f_{\rm min})^\gamma\right)^{-1}+P_{\text{white}}\;,
\end{equation}
where $P_{\text{white}}$ is the high-frequency, white-noise level, and
$P_0=P_{\text{white}}\,\left(f_{\rm knee}/f_{\rm min})^\gamma-1\right)$.  The
value of $\gamma$ is $\approx 1.7$ for LFI and 2 for HFI detectors.

A noise time-stream with this spectrum could be obtained by modelling it
in Fourier space with random coefficients and then performing an
FFT. However, considering the fact that the coherence length for an
HFI detector would be approximately $10^8$ samples,this method is
rather expensive in terms of main memory consumption.

The noise is more efficiently generated by sampling and adding
numerical solutions of stochastic differential equations (SDEs) of the
form
\begin{equation}
\label{eq:sde}
  \dot{x_i}(t) = -x_i(t)/\tau_i + \xi_i(t)\;,
\end{equation}
where $\tau_i$  is the autocorrelation time of  process $x_i(t)$.  The
$\xi_i(t)$ are individual white  noise processes which are implemented
in the discretised form  of Eq.~(\ref{eq:sde}) as normalised series of
independent  Gaussian random  numbers $\xi_i(t)$  with autocorrelation
function $\langle \xi_i(t)\,\xi_i(s) \rangle = \delta(t-s)$.

In order to have a good logarithmic frequency coverage, a logarithmic
grid of $\tau$ between $2\,\pi/f_{\rm knee}$ and $2\,\pi/f_{\rm min}$
is used together with a process with $\tau_0\rightarrow0$ for the
high-frequency white noise $n_0(t)$.

The resulting SDE spectra
\begin{equation}
  P_i(f) = \left(\tau_i^{-2} + (2\,\pi\,f)^2\right)^{-1}
\end{equation}
can be combined to an approximation of the target spectrum $P_n^*(f)$
(Eq.~\ref{eq:targetnoise}) by combining the individual processes with
properly chosen weights $c_i$:
\begin{eqnarray}
  n(t) &=& \sum_i\, c_i\, x_i(t)\,,\\
  P_n(f) &=& \sum_i\, c_i^2\, P_i(f) \approx P_n^*(f)\,.
\end{eqnarray}
More technical detail on choosing the weights, initialising the SDE
processes, and optimising the computational performance will be
covered in a separate publication (En{\ss}lin et al., in
preparation), but see also \cite{keihanen-etal-2004}.

Since the output of the noise generator is added to the output of the
{\tt sampler} module, both are naturally produced at the same sampling
frequency.

Parallelisation of the algorithm is not straightforward, because every
noise sample can only be computed after all of its predecessors have
been calculated.  This prohibits the most straightforward strategy of
dividing the requested samples between multiple processors. In our
approach, each SDE process calculates its associated time stream
separately (which can be done in parallel processes), and all of these
time streams are added at the end. While this solution can be
significantly faster than a scalar implementation, its scalability is
limited by the number of SDE processes ($\approx\!10$). Increasing the
number of processors beyond this value achieves no further speedup.

\section {Summary and outlook}
\label{summary}

We have described in this paper the current state of the \emph{Planck}
simulation pipeline. This package allows a complete simulation of
the time-ordered data streams produced by the satellite's detectors, taking as input
a set of cosmological parameters, various foreground emission models, a satellite
scanning strategy and parameters characterising the detectors and on-board
electronics. The availability of such data sets allows extensive testing of the data
analysis software for the mission already before launch, which greatly decreases the time
interval between the availability of raw data and the publication of first results.
Furthermore the simulation modules can be conveniently used to study the influence
of assumed systematic effects or changes in the mission parameters on the quality of the
final data.

Obviously, many of the package's modules are still
under development, and additions of new modules are expected. The
integration with the \emph{Planck} Data Management Component and the
mission's \emph{Integrated Data and Information System} (IDIS) will be
another important milestone. Any significant changes and enhancements
will be presented in a follow-up paper.

The non-\emph{Planck}-specific modules will hopefully be of use outside the \emph{Planck}
collaboration as well. This covers almost all of the presented codes,
with the exception of the modules for the satellite dynamics and the
sorption-cooler noise. A future public release of the non-\emph{Planck}-specific parts
of the package is envisaged by the authors. Currently, a partial simulation
pipeline, which is hosted by the German Astrophysical Virtual Observatory, can be tested
interactively over the Internet\footnote{\href{http://www.g-vo.org/portal/tile/products/services/planck/index.jsp}{http://www.g-vo.org/portal/tile/products/services/planck/index.jsp}}.

\begin{acknowledgements}

The work presented in this article is done in the course of the
\emph{Planck} satellite mission. The authors are grateful for many
productive interactions with the \emph{Planck} team.

We thank Mark Ashdown for implementation and support of the {\tt Beam} package;
Benjamin Wandelt for providing an initial implementation of
the {\tt oofnoise} and {\tt totalconvolver} modules;
Carlo Burigana, Davide Maino and Krzysztof G\'orski for early, FFT-based
noise generation modules;
Daniel Mortlock and Floor van Leeuwen for the {\tt simmission} module;
Ian Grivell and Bob Mann for the original sampler module;
Aniello Mennella and Michele Maris for the {\tt glissando} package;
Fabrizio Villa and Maura Sandri for instrumental information on LFI;
Bob Mann for compiling the first focal plane database and Mark Ashdown
for maintaining it;
Giovanna Giardino, Bj\o rn Sch\"afer, Christoph Pfrommer, Nabila Aghanim and
Fran\c cois Bouchet for the contribution of template maps;
Carlo Baccigalupi for contributing his dust emission model;
and Antony Lewis for helping with the integration of {\tt CAMB}.

Some of the codes presented in this paper are based on the HEALPix
package \citep{gorski-etal-2005}, whose usage is gratefully acknowledged.

\end{acknowledgements}

\bibliographystyle{aa}
\bibliography{planck}

\begin{thebibliography}{35}
\expandafter\ifx\csname natexlab\endcsname\relax\def\natexlab#1{#1}\fi

\bibitem[{{Bluestein}(1968)}]{bluestein-1968}
{Bluestein}, L.~I. 1968, Northeast Electronics Research and Engineering Meeting
  Record, 10, 218

\bibitem[{Box \& Muller(1958)}]{box-muller-1958}
Box, G.~E.~P. \& Muller, M.~E. 1958, Ann.\ Math.\ Stat., 29, 610

\bibitem[{{Challinor} {et~al.}(2000){Challinor}, {Fosalba}, {Mortlock},
  {Ashdown}, {Wandelt}, \& {G{\' o}rski}}]{challinor-etal-2000}
{Challinor}, A., {Fosalba}, P., {Mortlock}, D., {et~al.} 2000, \prd, 62, 123002

\bibitem[{{Challinor} {et~al.}(2002){Challinor}, {Mortlock}, {van Leeuwen},
  {Lasenby}, {Hobson}, {Ashdown}, \& {Efstathiou}}]{challinor-etal-2002}
{Challinor}, A.~D., {Mortlock}, D.~J., {van Leeuwen}, F., {et~al.} 2002,
  \mnras, 331, 994

\bibitem[{{Colberg} {et~al.}(2000){Colberg}, {White}, {Yoshida}, {MacFarland},
  {Jenkins}, {Frenk}, {Pearce}, {Evrard}, {Couchman}, {Efstathiou}, {Peacock},
  {Thomas}, \& {The Virgo Consortium}}]{colberg-etal-2000}
{Colberg}, J.~M., {White}, S.~D.~M., {Yoshida}, N., {et~al.} 2000, \mnras, 319,
  209

\bibitem[{{Dame} {et~al.}(1996){Dame}, {Hartmann}, \&
  {Thaddeus}}]{dame-etal-1996}
{Dame}, T.~M., {Hartmann}, D., \& {Thaddeus}, P. 1996, Bulletin of the American
  Astronomical Society, 28, 1362

\bibitem[{{Dame} {et~al.}(2001){Dame}, {Hartmann}, \&
  {Thaddeus}}]{dame-etal-2001}
---. 2001, \apj, 547, 792

\bibitem[{{Dolag} {et~al.}(2005){Dolag}, {Hansen}, {Roncarelli}, \&
  {Moscardini}}]{dolag-etal-2005}
{Dolag}, K., {Hansen}, F.~K., {Roncarelli}, M., \& {Moscardini}, L. 2005, ArXiv
  Astrophysics e-prints, submitted to \mnras

\bibitem[{{Finkbeiner}(2003)}]{finkbeiner-2003}
{Finkbeiner}, D.~P. 2003, \apjs, 146, 407

\bibitem[{{Finkbeiner} {et~al.}(1999){Finkbeiner}, {Davis}, \&
  {Schlegel}}]{finkbeiner-etal-1999}
{Finkbeiner}, D.~P., {Davis}, M., \& {Schlegel}, D.~J. 1999, \apj, 524, 867

\bibitem[{Frigo \& Johnson(1998)}]{frigo-johnson-1998}
Frigo, M. \& Johnson, S.~G. 1998, in Proc. 1998 IEEE Intl. Conf. Acoustics
  Speech and Signal Processing, Vol.~3 (IEEE), 1381--1384

\bibitem[{{G{\' o}rski} {et~al.}(2002){G{\' o}rski}, {Banday}, {Hivon}, \&
  {Wandelt}}]{gorski-etal-2002}
{G{\' o}rski}, K.~M., {Banday}, A.~J., {Hivon}, E., \& {Wandelt}, B.~D. 2002,
  in ASP Conf. Ser. 281: Astronomical Data Analysis Software and Systems XI,
  107

\bibitem[{{G{\' o}rski} {et~al.}(2005){G{\' o}rski}, {Hivon}, {Banday},
  {Wandelt}, {Hansen}, {Reinecke}, \& {Bartelmann}}]{gorski-etal-2005}
{G{\' o}rski}, K.~M., {Hivon}, E., {Banday}, A.~J., {et~al.} 2005, \apj, 622,
  759

\bibitem[{{Giardino} {et~al.}(2002){Giardino}, {Banday}, {G{\' o}rski},
  {Bennett}, {Jonas}, \& {Tauber}}]{giardino-etal-2002}
{Giardino}, G., {Banday}, A.~J., {G{\' o}rski}, K.~M., {et~al.} 2002, \aap,
  387, 82

\bibitem[{Grivell \& Mann(1999)}]{grivell-mann-1999}
Grivell, I. \& Mann, R. 1999, Sampler module for scan strategy simulations,
  {P}lanck LiveLink repository

\bibitem[{{Jenkins} {et~al.}(2001){Jenkins}, {Frenk}, {White}, {Colberg},
  {Cole}, {Evrard}, {Couchman}, \& {Yoshida}}]{jenkins-etal-2001}
{Jenkins}, A., {Frenk}, C.~S., {White}, S.~D.~M., {et~al.} 2001, \mnras, 321,
  372

\bibitem[{{Keih{\"a}nen} {et~al.}(2004){Keih{\"a}nen}, {Kurki-Suonio}, \&
  {Poutanen}}]{keihanen-etal-2004}
{Keih{\"a}nen}, E., {Kurki-Suonio}, H., \& {Poutanen}, T. 2004, ArXiv
  Astrophysics e-prints, astro-ph/0412517

\bibitem[{{Lewis} {et~al.}(2000){Lewis}, {Challinor}, \&
  {Lasenby}}]{lewis-etal-2000}
{Lewis}, A., {Challinor}, A., \& {Lasenby}, A. 2000, \apj, 538, 473

\bibitem[{Marsaglia(2003)}]{marsaglia-2003}
Marsaglia, G. 2003, Journal of Statistical Software, 8

\bibitem[{{Neugebauer} {et~al.}(1971){Neugebauer}, {Miinch}, {Kieffer},
  {Chase}, \& {Miner}}]{neugebauer-etal-1971}
{Neugebauer}, G., {Miinch}, G., {Kieffer}, H., {Chase}, S.~C., \& {Miner}, E.
  1971, \aj, 76, 719

\bibitem[{NOST(1999)}]{fits-1999}
NOST. 1999, Definition of the Flexible Image Transport System (FITS),
  http://archive.stsci.edu/fits/fits\_standard/

\bibitem[{{Pasian} \& {Sygnet}(2002)}]{pasian-sygnet-2002}
{Pasian}, F. \& {Sygnet}, J. 2002, in Astronomical Data Analysis II. Edited by
  Starck, Jean-Luc; Murtagh, Fionn D. Proceedings of the SPIE, Volume 4847.,
  25--34

\bibitem[{{Pence}(1999)}]{pence-1999}
{Pence}, W. 1999, in ASP Conf. Ser. 172: Astronomical Data Analysis Software
  and Systems VIII, 487

\bibitem[{{Sch{\"a}fer} {et~al.}(2004{\natexlab{a}}){Sch{\"a}fer}, {Pfrommer},
  {Bartelmann}, {Springel}, \& {Hernquist}}]{schaefer-etal-2004a}
{Sch{\"a}fer}, B.~M., {Pfrommer}, C., {Bartelmann}, M., {Springel}, V., \&
  {Hernquist}, L. 2004{\natexlab{a}}, ArXiv Astrophysics e-prints,
  astro-ph/0407089, submitted to \mnras

\bibitem[{{Sch{\"a}fer} {et~al.}(2004{\natexlab{b}}){Sch{\"a}fer}, {Pfrommer},
  {Hell}, \& {Bartelmann}}]{schaefer-etal-2004b}
{Sch{\"a}fer}, B.~M., {Pfrommer}, C., {Hell}, R., \& {Bartelmann}, M.
  2004{\natexlab{b}}, ArXiv Astrophysics e-prints, astro-ph/0407090, submitted
  to \mnras

\bibitem[{{Schlegel} {et~al.}(1998){Schlegel}, {Finkbeiner}, \&
  {Davis}}]{schlegel-etal-1998}
{Schlegel}, D.~J., {Finkbeiner}, D.~P., \& {Davis}, M. 1998, \apj, 500, 525

\bibitem[{{Seljak} \& {Zaldarriaga}(1996)}]{seljak-zaldarriaga-1996}
{Seljak}, U. \& {Zaldarriaga}, M. 1996, \apj, 469, 437

\bibitem[{{Swarztrauber}(1982)}]{swarztrauber-1982}
{Swarztrauber}, P. 1982, {Vectorizing the Fast Fourier Transforms} (New York:
  Academic Press), 51

\bibitem[{Taffoni {et~al.}(2005)Taffoni, Maino, Vuerli, Castelli, Smareglia,
  Zacchei, En{\ss}lin, \& Pasian}]{taffoni-etal-2005}
Taffoni, G., Maino, D., Vuerli, C., {et~al.} 2005, IEEE Transactions on
  Computers, in press

\bibitem[{{Tauber}(2004)}]{tauber-2004}
{Tauber}, J.~A. 2004, Advances in Space Research, 34, 491

\bibitem[{{Valls-Gabaud}(1998)}]{valls-gabaud-1998}
{Valls-Gabaud}, D. 1998, PASA, 15, 111

\bibitem[{{van Leeuwen} {et~al.}(2002){van Leeuwen}, {Challinor}, {Mortlock},
  {Ashdown}, {Hobson}, {Lasenby}, {Efstathiou}, {Shellard}, {Munshi}, \&
  {Stolyarov}}]{vanLeeuwen-etal-2002}
{van Leeuwen}, F., {Challinor}, A.~D., {Mortlock}, D.~J., {et~al.} 2002,
  \mnras, 331, 975

\bibitem[{{Wandelt} \& {G{\' o}rski}(2001)}]{wandelt-gorski-2001}
{Wandelt}, B.~D. \& {G{\' o}rski}, K.~M. 2001, \prd, 63, 123002

\bibitem[{{Wright}(1976)}]{wright-1976}
{Wright}, E.~L. 1976, \apj, 210, 250

\bibitem[{{Zaldarriaga} \& {Seljak}(1997)}]{zaldarriaga-seljak-1997}
{Zaldarriaga}, M. \& {Seljak}, U. 1997, \prd, 55, 1830

\end{thebibliography}
\end{document}